\documentclass[11pt]{article}

\usepackage{amsmath}
\usepackage{amsfonts}
\usepackage{amssymb}
\usepackage[english]{babel}
\usepackage[colorinlistoftodos]{todonotes}
\usepackage{comment}  
\usepackage{float}
\usepackage{tikz} 
\usepackage{xparse}
\usepackage{enumerate}
\usepackage{enumitem}
\usepackage{blindtext}
\usepackage[T1]{fontenc}
\usepackage{microtype}
\usepackage{graphicx}
\usepackage{booktabs} 
\usepackage{algorithmic}
\usepackage{algorithm}
\usepackage{times}
\usepackage{epsfig}
\usepackage{amsthm}
\usepackage[font=small,labelfont=bf]{caption}

\usepackage{array}
\newcolumntype{P}[1]{>{\hspace{0pt}}p{#1}}

\usepackage[export]{adjustbox}
\usepackage{rotating}
\usepackage{csvsimple,longtable,booktabs}
\usepackage{tabularx}
\usepackage[section]{placeins}

\usepackage{hyperref}
\usepackage{authblk}

\newcommand{\bigos}{\mathcal{O}^*}

\newtheorem{remark}{Remark}
\newtheorem{theorem}{Theorem}

\begin{document}

\title{An Improved Fixed-Parameter Algorithm for 2-Club Cluster Edge Deletion}

\author{Faisal N. Abu-Khzam}
\author{Norma Makarem}
\author{Maryam Shehab}

\affil{Department of Computer Science and Mathematics\\ 
Lebanese American University\\
Beirut, Lebanon}
 
\date{}
\maketitle

\begin{abstract}
A 2-club is a graph of diameter at most two. In the decision version of the parametrized {\sc 2-Club Cluster Edge Deletion} problem, an undirected graph $G$ is given along with an integer $k\geq 0$ as parameter, and the question is whether $G$ can be transformed into a disjoint union of 2-clubs by deleting at most $k$ edges. A simple fixed-parameter algorithm solves the problem in $\bigos(3^k)$, and a decade-old algorithm was claimed to have an improved running time 
of $\bigos(2.74^k)$ via a sophisticated case analysis. 
Unfortunately, this latter algorithm suffers from a flawed branching scenario. In this paper, an improved fixed-parameter algorithm is presented with a running time in $\bigos(2.695^k)$. 
\end{abstract}


\thispagestyle{empty}

\section{Introduction}

A graph modification problem typically requires some minimal number of operations, referred to as graph editing, to transform a given graph into one that has a desired property, or structure. When restricted to edge editing operations, namely the addition or deletion of an edge, the practical objective is to make ``corrections'' to the graph by eliminating false positives (edge removal) and/or false negatives (edge addition). 
If edge deletion only is required, the objective can also be to partition the vertex set of a graph into subsets that satisfy the desired property. 

A typical popular problem in this area is {\sc Cluster Editing}, which is known as a model for correlation clustering. The problem seeks a transformation of an input graph into a disjoint union of cliques via a user-specified (or minimum) number of edge editing operations. Cluster Editing received a notable attention in the parameterized complexity literature  \cite{abu17,abu18,Bocker13,Boecker2011,DBLP:conf/ciac/GrammGHN03,DBLP:journals/algorithmica/GrammGHN04,guo2009more,HeggernesLNPT10,KU12,SHAMIR2004173}, and it has found application in various practical settings \cite{barr9030927,Barr-IJSC2020,barrTransAI,DAddario2014,dehne2006cluster,fadiel2006computational,Huffner2010}.
In various application scenarios, the requirement for clusters to be cliques is found to be too restrictive; hence, some relaxed clique models for dense subgraph have been proposed as alternatives. Examples include quasi-clique, $s$-plex and $s$-club \cite{Lee2010}. In this paper we merely consider the notion of a 2-club, being a natural extension of a clique, or 1-club, and also because in social networks nodes that are at distance two from each other are often expected to be closely related \cite{Liben-NowellK07}.

Many variants of {\em editing} a graph into a disjoint union of 2-clubs have been studied, such as {\sc 2-Club Cluster Vertex Deletion}, {\sc 2-Club Cluster Edge Deletion}, and {\sc 2-Club Cluster Editing}. All these variants are ${\cal NP}$-Complete \cite{Liu12}. 
Moreover, it was shown in \cite{FigielHNN21} that {\sc 2-Club Cluster Editing} is ${\cal W}$[2]-hard with respect to the number of modified edges, hence most likely not fixed-parameter tractable ($FPT$). In addition, the {\sc 2-Club Cluster Vertex Deletion} version of the problem was shown to be $FPT$ but not poly-kernelizable (unless $NP\subset$ co-$NP/poly$.) Moreover, the problem was shown not to have a subexponential-time algorithm modulo the Exponential-Time Hypothesis \cite{MisraPS20}.

In this paper we are mainly interested in the {\sc 2-Club Cluster Edge Deletion} problem, which we believe is a natural extension of {\sc Cluster Editing} being a possibly-better model for correlation clustering. In \cite{Liu12}, Liu et al. presented a fixed-parameter algorithm for the problem, with a running time that was claimed to be in $\bigos(2.74^k)$. Unfortunately, the claimed asymptotic running time was based on a branching scenario that omitted a critical case. We shall provide a brief note about the flawed argument in the appendix.
 
\section{Preliminaries}

We consider simple undirected unweighted graphs, and we use common graph theoretic terminology such as those found in \cite{West}. 
Let $G=(V,E)$ be a simple undirected unweighted graph. The distance between two vertices $u$ and $v$ in $G$, denoted $d(u,v)$, is the length of a shortest path between them. The diameter of a connected graph $G$ is the maximum distance between any two vertices. 

For a vertex $v \in V$, the set of vertices at distance $t$ from $v$ is denoted by $N_t(v)$, and the set of all vertices that are at distance at most $t$ from $v$ is denoted by $N_t[v]$. In particular the open and closed neighborhoods of $v$ are, respectively, $N(v) = N_1(v)=\{w\in V: uw\in E\}$ and $N[v]= N(v)\cup\{v\}$. Since we are dealing with simple graphs (with no multiple edges or self loops), the degree of a vertex $v$ is $degree(v)=|N(v)|$. A vertex of degree one is referred to as a pendant vertex.

A simple path $P$ in $G$ is an ordered sequence of pairwise distinct vertices $(v_1, v_2,  \ldots v_{k})$ such that $v_iv_{i+1}\in E$ for all $i \in \{1,\ldots k\}$. $P$ is an induced path if these are the only edges between its vertices. The length of $P$ is $k-1$ in this case and a path of length $t$ is denoted by $P_t$ (so we assume the number of vertices in $P_t$ is $t+1$). 
A tail of length $k$, or $k$-tail, is an induced path with degree-two internal vertices and with one endpoint that is of degree one in $G$. A 3-tail is shown in Figure \ref{tail} (next section).  

A clique in a graph $G$ is a set of pair-wise adjacent vertices. An $s$-club is a set of vertices any two of which are at distance at most $s$ from each other. As such, a clique is nothing but a 1-club.
As mentioned in the previous section, the main contribution of this paper is an improved fixed-parameter algorithm for the 2-Clubs Edge Deletion problem, which we formally define as follows.

\vspace{10pt}
\noindent
{\sc 2-Club Cluster Edge Deletion (2CCED)}

\noindent
\underline{Given:} a graph $G$ and an integer $k$

\noindent
\underline{Question:} can $G$ be transformed into a disjoint union of 2-clubs by deleting at most $k$ edges?

\vspace{10pt}

The 2CCED problem is {\cal NP}-Complete, as shown in \cite{Liu12}. However, the hardness proof does not work for bounded-degree graphs, which can be of special importance since any 2-club is of bounded size in this case. 
Observe that 2CCED is trivially solvable in polynomial time when the maximum degree is bounded above by two: if a connected component of the graph is a path $P = (v_1, v_2,\ldots v_s)$, we simply successively delete edges $v_{3i}v_{3i+1}$ for $i = 1,2\ldots$, which is optimum in this case. On the other hand, if a connected component is a cycle of length $> 5$ then we delete an arbitrary edge and the resulting graph will be an isolated path that can be resolved as discussed. 

A solution to the 2CCED problem yields a graph whose connected components are diameter-two subgraphs. We refer to the resulting graph as a 2-clubs graph. The presence of a path of length three whose endpoints are at distance exactly three from each other is the main ``forbidden structure'' that prevents a graph from being a 2-clubs graph. We shall refer to such a path as a {\em conflict quadruple} in this paper.
During the search for a solution we look for a conflict quadruple and try to {\em resolve} it by deleting one of the three edges forming it. We shall mark some edges as permanent if we decide they are not to be part of a solution (hence not to be deleted).

\section{An Improved 2CCED Algorithm}

Our algorithm is simply based on resolving any conflict quadruple by deleting one of the three edges forming it. In each case (or branch) the parameter $k$ is decreased by one. This general approach gives a simple $\bigos(3^k)$ algorithm. However, there are cases where more than one conflict intersect in a way that allows us to further reduce the parameter at some branches. Moreover, there are simpler cases where we know exactly which edge (or group of edges) to delete ``without loss of optimality.'' Such cases can be dealt with as part of a polynomial-time procedure that is based on reduction rules. 

\subsection{A Reduction Procedure}

A reduction procedure is assumed to be exhaustively applied before the search-tree backtracking algorithm and during the search process, prior to any choice, or decision, made by the search algorithm. 
The main reduction rules are given below. They are assumed to be applied successively in such a way that a rule is not applied, until all the previous rules have been applied exhaustively. We shall prove the soundness of non-obvious reduction rules only.

\vspace{5pt}
\noindent
\textbf{Reduction Rule 1.} The algorithm terminates and reports a no instance whenever the parameter $k$ becomes negative.

\vspace{5pt}
\noindent
\textbf{Reduction Rule 2.}
The algorithm terminates and reports a yes instance if the graph becomes empty (assuming $k\geq 0$ due to the previous rule).

\vspace{5pt}
\noindent
\textbf{Reduction Rule 3.} If $G$ contains a connected component $C$ that is a 2-club, then delete $C$. 

\vspace{5pt}
\noindent
Note that exhaustive application of Rule 3 results also in deleting all isolated vertices.

\vspace{5pt}
\noindent
\textbf{Reduction Rule 4.} If two non-adjacent vertices $a$ and $b$ have more than $k$ common neighbours then delete the edges linking $a$ and $b$ to $N(a)\setminus N_2[b]$ and $N(b)\setminus N_2[a]$ respectively.

\vspace{5pt}
\noindent
\textbf{Soundness.} Since $a$ and $b$ have more than $k$ common neighbors it would be impossible to cause the distance between them to increase beyond two, so they must belong to the same 2-club, which does not contain elements of 
$N(a)\setminus N_2[b]\cup N(b)\setminus N_2[a]$.

\vspace{5pt}
\noindent
\textbf{Reduction Rule 5.} If $G$ contains a connected component $H$ of maximum degree two, then $H$ can be transformed optimally into a 2-clubs (sub)graph. This results in decreasing the value of $k$ by the number of deleted edges.

\vspace{5pt}
\noindent
\textbf{Soundness.}
Any connected component of maximum degree two is either a cycle or a path, which can be resolved as described in Section 2 above.

\begin{figure}[htb!]
    \centering
\begin{tikzpicture}[main/.style = {draw, circle, node distance=1.5cm, scale = 1, minimum size=2em}] 
\node[main] (1) {a};
\node[main] (2) [right of=1] {b};
\node[main] (3) [right of=2] {c};
\node[main] (4) [right of=3] {d};
\node[main] (5) [above left=of 1] {x};
\node[main] (6) [below left=of 1] {y};
\draw (1) -- (2);
\draw (2) -- (3);
\draw (3) -- (4);
\draw (1) -- (5);
\draw (1) -- (6);
\end{tikzpicture}
    \caption{A tail of length three.}
    \label{tail}
\end{figure}
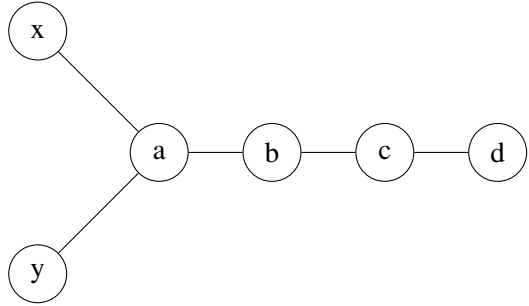

\vspace{5pt}
\noindent
\textbf{Reduction Rule 6.} If $G$ has a 3-tail $T=(a,b,c,d)$, as in Figure \ref{tail}, then we simply delete the edge $ab$ and decrease $k$ by 1.

\vspace{5pt}
\noindent
\textbf{Soundness.}
Since the two vertices $a$ and $d$ must belong to two different 2-clubs, at least one of the three edges forming $T$ must be deleted. Deleting $ab$ results in an isolated 2-club (namely the path formed by $b,c$ and $d$) and cannot result in a sub-optimal solution.

\vspace{10pt}

\subsection{Branching Rules}

We now present our bounded search tree algorithm, which simply works in a recursive manner and can be viewed as a search-tree traversal. 
The running time is thus proportional (modulo a polynomial factor) to the number of recursive calls. This is why we use the $\bigos$ notation, which mainly displays the total number of recursive calls and hides any polynomial factor.

In what follows, we consider an instance $(G,k)$ of 2CCED that has been pre-processed by exhaustive application of the reduction rules. As mentioned earlier, the reduction rules are assumed to be applied exhaustively whenever they are applicable during the search process. As such, we either have a solution (when $G$ becomes empty) or every connected component of $G$ contains at least one vertex of degree $\geq 3$ and at least two vertices that are at distance exactly three from each other. This order of events applies also to the branching rules, given by a list of cases below. Therefore, in each case, we assume none of the previously addressed conditions hold. 

\subsubsection*{Case 1. Neighbors of endpoints of a $P_2$.}

If we have an induced path of length two, say $P=(a,b,c)$, such that $|N(a)\setminus N_2[c] \cup N(c)\setminus N_2[a]| \geq 2$, then we branch by either (i) deleting $ab$ or $bc$ or all the vertices in $N(a)\setminus N_2[c] \cup N(c)\setminus N_2[a]$. The worst-case recurrence is thus $T(k) = 2T(k-1) + T(k-2)$ with a corresponding running time in 
$\bigos(2.415^k)$.

\vspace{5pt}
\noindent
{\bf Soundness.}
Each of the first two branches deletes one of the three edges of a conflict quadruple that contains $(a,b,c)$ as a sub-path. In the third case (or branch) the two edges $ab$ and $bc$ become permanent. Thus any neighbor of $a$ that is at distance three from $c$ must be deleted, and vice versa.

\begin{remark}
The above branching scenario applies implicitly in two notable cases that we shall (therefore) exclude in the sequel.
\begin{itemize}
   
\item[-] If we have a conflict quadruple $(a,b,c,d)$ with degree-two internal vertices ($b$ and $c$), then any neighbor of $a$ is at distance exactly three from $c$, and the same applies to $d$ and $b$. Thus the path $(a,b,c)$ satisfies the branching condition of Case 1, so from this point on this case is implicitly excluded.
   
\item[-] If we have a pair of vertices $u$ and $v$ that are at distance four from each other, then the three internal vertices on a shortest path between $u$ and $v$ also satisfy the condition of Case 1.
\end{itemize}

\end{remark}

Based on the above remark, we can assume that from this point on every connected component of $G$ is a 3-club. 
Moreover, any such 3-club contains at least one vertex $a$ with a non-empty $N_3(a)$ and every vertex in $N_2(a)$ has at most one neighbor in $N_3(a)$ (if an element of $N_2(a)$ has two or more neighbors in $N_3(a)$ then Case 1 would be applicable). 

In the following cases and sub-cases we assume we have a conflict quadruple $P = (a,b,c,d)$, and we mainly seek to resolve it by deleting one of the three edges. In some cases, we might also consider other conflict quadruples, if found in the neighborhood of $P$. 

\subsubsection*{Case 2. Conflict quadruple with pendant endpoints.} In the special case where every conflict quadruple $(a,b,c,d)$ satisfies $degree(a)=degree(b)=1$, we know the internal vertices $b$ and $c$ do not have more than one degree-one neighbor, otherwise Case 1 applies. Therefore deleting the edge $bc$ can only resolve exactly one conflict and it could possibly yield more conflicts, while deleting $ab$ or $cd$ can resolve one or more conflicts without leading to more conflict quadruples. Therefore in this special case we simply branch by either deleting $ab$ or $cd$, with a corresponding running time in $\bigos(2^k)$.

\vspace{5pt}
From this point on, and without loss of generality, we shall assume $d$ has at least one neighbor other than $c$. Such a neighbor is therefore at distance one or two from $b$. In fact, if its distance to $b$ is three, then Case 1 would apply to the path $(b,c,d)$. 

\subsubsection*{Case 3. $b$ has a neighbor at distance one from $d$.} 
\begin{figure}[htb!]
    \centering
\begin{tikzpicture}[main/.style = {draw, circle, node distance=1.5cm, scale = 1.2, minimum size=2em},
every edge quotes/.style = {auto=left, sloped, font=\scriptsize, inner sep=1pt}
] 
\node[main] (1) {a};
\node[main] (2) [right of=1] {b};
\node[main] (3) [right of=2] {c};
\node[main] (4) [right of=3] {d};
\node[main] (5) [below of=3] {w};
\draw (1) -- node[above, sloped, pos=0.5] {} (2);
\draw (2) -- node[above, sloped, pos=0.5] {} (3);
\draw (3) -- node[above, sloped, pos=0.5] {} (4);
\draw (2) -- node[below, sloped, pos=0.5] {} (5);
\draw (4) -- node[below, sloped, pos=0.5] {} (5);
\end{tikzpicture}
    \caption{Induced paths with a common edge.}
    \label{common1}
\end{figure}
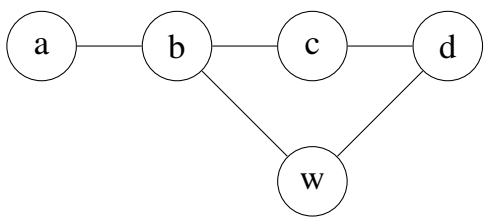

Let $w$ be a common neighbor of $b$ and $d$, as shown in Figure \ref{common1}. We branch as follows: 

\begin{itemize}

\item[-] delete edge $ab$; 

\item[-] delete edges $bc$ and $bw$;
    
\item[-] delete edges $cd$ and $dw$; 

\item[-] delete edges $bc$ and $dw$; 

\item[-] delete edges $cd$ and $bw$.
\end{itemize}
 
This gives the recurrence: $T(k)= T(k-1) + 4T(k-2)$ with a corresponding running time in $\bigos(2.562^k)$. 

\vspace{10pt}

\vspace{5pt}
\noindent
{\bf Soundness.}
After the second branch, we know that $b$ and at least one vertex from the pair $\{c,w\}$ is in the same 2-club as $a$. If both $c$ and $w$ are in this 2-club, then we must delete $cd$ and $dw$ (since $d$ cannot be in the same club). This justifies the third branch. After the third branch, the 2-club of $a$ contains either $\{a,b,w\}$ so we delete $bc$ and $dw$, or it contains $\{a,b,c\}$ and this leads to deleting $bw$ and $cd$. \qed

\begin{remark}
Observe that not all links are shown in the above figure, but the branching scenario can only be improved if other links exist without affecting the distance between $a$ and $d$. For example, adding an edge between $c$ and $w$ leads to a better recurrence since $cw$ would have to be deleted in each of the last two branches.
\end{remark}

\subsubsection*{Case 4. $b$ has a neighbor at distance two from $d$.}
\begin{figure}[htb!]
    \centering
\begin{tikzpicture}[main/.style = {draw, circle, node distance=1.5cm, scale = 1.2, minimum size=2em},
every edge quotes/.style = {auto=left, sloped, font=\scriptsize, inner sep=1pt}
] 
\node[main] (1) {a};
\node[main] (2) [right of=1] {b};
\node[main] (3) [right of=2] {c};
\node[main] (4) [right of=3] {d};
\node[main] (5) [below of=3] {x};
\node[main] (6) [below of=4] {y};
\draw (1) -- node[above, sloped, pos=0.5] {} (2);
\draw (2) -- node[above, sloped, pos=0.5] {} (3);
\draw (3) -- node[above, sloped, pos=0.5] {} (4);
\draw (2) -- node[below, sloped, pos=0.5] {} (5);
\draw (4) -- node[below, sloped, pos=0.5] {} (6);
\draw (6) -- node[below, sloped, pos=0.5] {} (5);

\end{tikzpicture}
    \caption{Two vertices at distance three from a given vertex.}
    \label{2.2}
\end{figure}
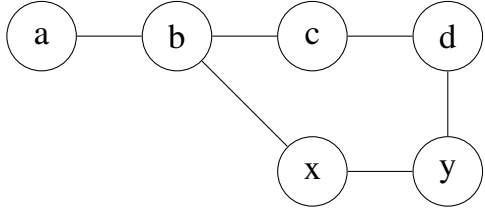

Let $(b,x,y,d)$ be an induced $P_3$ corresponding to this case, as shown in Figure \ref{2.2}.

\vspace{10pt}

The distance between $a$ and $y$ in the above figure leads to two possible sub-cases, namely $d(a,y)=3$ and $d(a,y)=2$. 

\subsubsection*{Case 4.1. $d(a,y) = 3$.}
In this particular case we branch as follows:

\begin{itemize}

\item[-] delete $ab$;
\item[-] delete $bc$ and $bx$;
\item[-] delete $cd$ and $xy$;
\item[-] delete $bx$ and $cd$; 
\item[-] delete $bc$ and $xy$.

\end{itemize}

This again yields the recurrence $T(k) = T(k-1) + 4T(k-2)$ with a running time in
$\bigos(2.562^k)$.

\vspace{5pt}
\noindent
{\bf Soundness.}
After the second branch we are sure that $a$ is in the same club as $c$ or $x$ (or both). If $a,c$ and $x$ are in the same club, then we must delete edges $cd$ and $xy$, which corresponds to (and justifies) the third branch. Otherwise, we have exactly two cases: either $bx$ is deleted or $bc$ is deleted. In the first case, we must also delete $cd$ and in the second we must delete $xy$. \qed

\subsubsection*{Case 4.2. $d(a,y) = 2$.}
This is depicted in Figure \ref{4.2} below.
We further note that $d(v,c)$ is either 3 or 2 (if $d(v,c)=1$ then Case 3 would have been applied). If $d(v,c) = 3$, then the path $(a,b,c)$ would satisfy the condition of Case 1. Therefore we restrict our attention to the case where $d(v,c)=2$, and let $w$ be the common neighbor of $c$ and $v$. We further distinguish the two cases where $w\neq b$ and $w=b$.

\vspace{10pt}

\begin{figure}[htb!]
    \centering
\begin{tikzpicture}[main/.style = {draw, circle, node distance=1.5cm, scale = 1.2, minimum size=2em},
every edge quotes/.style = {auto=left, sloped, font=\scriptsize, inner sep=1pt}
] 
\node[main] (1) {a};
\node[main] (2) [right of=1] {b};
\node[main] (3) [right of=2] {c};
\node[main] (4) [right of=3] {d};
\node[main] (5) [below of=3] {x};
\node[main] (6) [below of=4] {y};
\node[main] (7) [below of=1] {v};
\draw (1) -- node[above, sloped, pos=0.5] {} (2);
\draw (1) -- node[above, sloped, pos=0.5] {} (7);
\draw (2) -- node[above, sloped, pos=0.5] {} (3);
\draw (3) -- node[above, sloped, pos=0.5] {} (4);
\draw (2) -- node[below, sloped, pos=0.5] {} (5);
\draw (4) -- node[below, sloped, pos=0.5] {} (6);
\draw (6) -- node[below, sloped, pos=0.5] {} (5);
\path (7) [bend right] edge node [above] {} (6);

\end{tikzpicture}
    \caption{Case 4.2.}
    \label{4.2}
\end{figure}
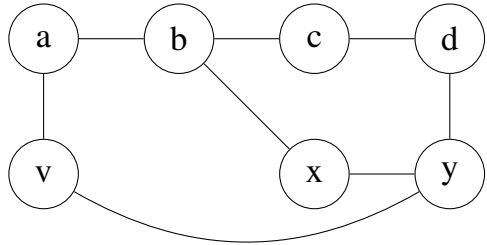

\subsubsection*{Case 4.2.1 $w \neq b$}

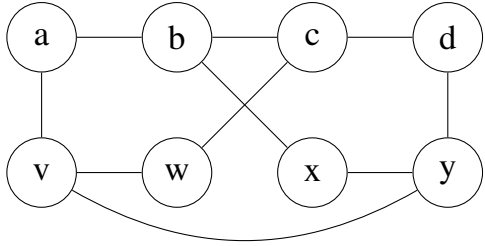
\begin{figure}[htb!]
    \centering
\begin{tikzpicture}[main/.style = {draw, circle, node distance=1.5cm, scale = 1.2, minimum size=2em},
every edge quotes/.style = {auto=left, sloped, font=\scriptsize, inner sep=1pt}
] 
\node[main] (1) {a};
\node[main] (2) [right of=1] {b};
\node[main] (3) [right of=2] {c};
\node[main] (4) [right of=3] {d};
\node[main] (5) [below of=3] {x};
\node[main] (6) [below of=4] {y};
\node[main] (7) [below of=1] {v};
\node[main] (8) [below of=2] {w};
\draw (1) -- node[above, sloped, pos=0.5] {} (2);
\draw (2) -- node[above, sloped, pos=0.5] {} (3);
\draw (3) -- node[above, sloped, pos=0.5] {} (4);
\draw (2) -- node[below, sloped, pos=0.5] {} (5);
\draw (4) -- node[below, sloped, pos=0.5] {} (6);
\draw (6) -- node[below, sloped, pos=0.5] {} (5);
\draw (1) -- node[above, sloped, pos=0.5] {} (7);
\draw (7) -- node[above, sloped, pos=0.5] {} (8);
\draw (3) -- node[above, sloped, pos=0.5] {} (8);
\path (7) [bend right] edge node [above] {} (6);

\end{tikzpicture}
    \caption{The case $b\neq w$}
    \label{alllinks}
\end{figure}

In this case we branch to resolve the conflict quadruple $(a,v,y,d)$ as follows (see Figure \ref{alllinks}):

\begin{itemize}
    \item[(1)] delete edge $dy$ and further branch to deleting
    \subitem $cd$
    \subitem $bc$
    \subitem $ab$, and $bx$ or $xy$ (to disconnect $d$ from $y$);
    
    \item[(2)] delete $av$ and further branch to deleting:
    \subitem $ab$
    \subitem $bc$, and $bx$ or $xy$
    \subitem $cd$, and $bx$ or $xy$;
    
    \item[(3)] delete $vy$ and further branch to deleting:
    
    \subitem $cd$, and $bx$ or $xy$
    
    \subitem $ab$, and $cw$ or $vw$ (since $d(a,c)=3$ after the deletion of $ab$)
    
    \subitem $bc$, $bx$, and $cw$ or $vw$ (to disconnect $c$ from $a$)
    
    \subitem $bc$, $xy$, and $cw$ or $vw$ (same reason).

\end{itemize}

\vspace{10pt}

\noindent
This gives the recurrence $T(k) = 3T(k-2) + 10T(k-3) + 4T(k-4)$ with a running time in
$\bigos(2.695^k)$.

\vspace{5pt}
\noindent
{\bf Soundness.}
We prove the soundness of each branching action separately. 

In the first branch we delete $dy$, being one of the edges of the conflict quadruple $(a,v,y,d)$, and proceed into resolving the conflict quadruple $(a,b,c,d)$. In this case, after the second (sub)branch we know $bc$ and $cd$ are permanent so we must delete $bx$ or $xy$ to make sure $d$ and $y$ are not in the same club (since we deleted of edge $dy$, which forces $d$ and $y$ to be in different 2-clubs).

In the second branch we proceed by deleting $av$ of $(a,v,y,d)$, and we know $dy$ is permanent. When we delete $bc$, the distance between $b$ and $d$ must become three. Otherwise, we would have a common neighbor between $b$ and $d$ other than $c$ and Case 3 would have been applied. Therefore we have another conflict quadruple to resolve, namely $(b,x,y,d)$. So we branch by deleting either $bx$ or $xy$ (since $dy$ is permanent in this branching case). The same applies to the sub-case (or sub-branch) where we delete $cd$ (since $d(b,d)$ becomes three again).

Finally, we note the importance of the order by which the quadruple $(a,b,c,d)$ is resolved in the third branch. First, the deletion of $cd$ again leads to $d(b,d)=3$ which is resolved by deleting $bx$ or $xy$. Second, the deletion of $ab$ increases the distance between $a$ and $c$ to three (same argument as in the case of $b$ and $d$). We thus have to resolve the conflict quadruple $(a,v,w,c)$ by deleting $cw$ or $vw$ (since $av$ is permanent in this branch). Finally, when deleting $bc$ we introduce two conflict quadruples: $(b,x,y,d)$ and $(a,v,w,c)$, which are resolved by deleting $bx$ or $xy$ and (in each case) deleting $cw$ or $vw$. \qed

\subsubsection*{Case 4.2.2 $w = b$} 

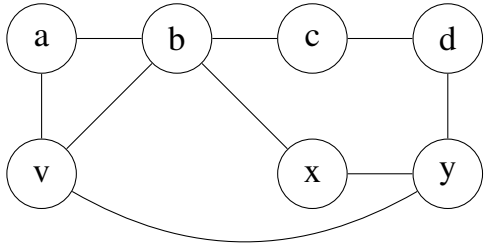
\begin{figure}[htb!]
    \centering
\begin{tikzpicture}[main/.style = {draw, circle, node distance=1.5cm, scale = 1.2, minimum size=2em},
every edge quotes/.style = {auto=left, sloped, font=\scriptsize, inner sep=1pt}
] 
\node[main] (1) {a};
\node[main] (2) [right of=1] {b};
\node[main] (3) [right of=2] {c};
\node[main] (4) [right of=3] {d};
\node[main] (5) [below of=3] {x};
\node[main] (6) [below of=4] {y};
\node[main] (7) [below of=1] {v};
\draw (1) -- node[above, sloped, pos=0.5] {} (2);
\draw (2) -- node[above, sloped, pos=0.5] {} (3);
\draw (3) -- node[above, sloped, pos=0.5] {} (4);
\draw (2) -- node[below, sloped, pos=0.5] {} (5);
\draw (4) -- node[below, sloped, pos=0.5] {} (6);
\draw (6) -- node[below, sloped, pos=0.5] {} (5);
\draw (1) -- node[above, sloped, pos=0.5] {} (7);
\draw (7) -- node[above, sloped, pos=0.5] {} (2);
\path (7) [bend right] edge node [above] {} (6);

\end{tikzpicture}
    \caption{The case $b = w$.}
    \label{w=b}
\end{figure}

In this case we also branch to resolve the conflict quadruple $(a,v,y,d)$ as follows (see Figure \ref{w=b}): 
\begin{itemize}
    \item[(1)] delete edge $dy$ and further branch to deleting
    \subitem $cd$
    \subitem $bc$
    \subitem $ab$, and $bx$ or $xy$ (to disconnect $d$ from $y$);
    
    \item[(2)] delete $av$ and further branch to deleting:
    \subitem $ab$ 
    \subitem $bc$, and $bx$ or $xy$
    \subitem $cd$, and $bx$ or $xy$;
    
    \item[(3)] delete $vy$ and further branch to deleting:
    
    \subitem $cd$, and $bx$ or $xy$
    
    \subitem $bc$, and $bx$ or $xy$
    
    \subitem $ab$ and $vb$ (to make sure $a$ is disconnected from $b$).
    
\end{itemize}

\vspace{10pt}

\noindent
This gives the recurrence $T(k) = 3T(k-2) + 11T(k-3)$ with a running time in
$\bigos(2.67^k)$.

\vspace{5pt}
\noindent
{\bf Soundness.}
The only difference between this case and the previous one is in the very last branch, when deleting $vy$ and $ab$. In this case we must make sure $a$ and $b$ are in different clubs (since we deleted $ab$), so we further delete $vb$ since $av$ is permanent in this last case. \qed

\vspace{5pt}
The above branching scenarios cover all the possible cases where we can find two vertices at distance three from each other in a graph that is not a disjoint union of 2-clubs. Therefore we can now state our main result.

\begin{theorem}
The 2-Club Cluster Edge Deletion problem is solvable in $\bigos(2.695^k)$. 
\end{theorem}

\section{Concluding Remarks}

We presented an improved fixed-parameter algorithm for {\sc 2-Club Cluster Edge Deletion}. 
The main approach is based on gradual elimination of favorable scenarios: bounded-degree-two, tail of length three, special paths of length two, paths of length four, etc... At each branching step, the absence of previous favorable scenarios makes it possible to improve the branching factor. Despite its practical importance, we believe the 
problem has not received enough attention, thus far. In fact, the only known FPT algorithm that improves on the exhaustive (folklore) $\bigos(3^k)$ method is the decade-old algorithm of Liu et al. \cite{Liu12}, which is shown to have a flawed branching case (as we prove in the appendix). 

The importance of the {\sc 2-Club Cluster Edge Deletion} problem stems from its ability to provide a better model for correlation clustering than the well studied {\sc Cluster Editing} problem. From a technical standpoint, the number of edge modifications (the parameter $k$) can be much smaller since the amount of edge additions needed to turn each resulting component into a clique can be very large. As such, correlation clustering via {\sc 2-Club Cluster Edge Deletion} can be more practical and possibly more informative. It would be interesting to have a fixed-parameter algorithm for the {\sc 3-Club Cluster Edge Deletion} problem using techniques similar to what we presented in this paper.
 
\bibliographystyle{abbrv}
\bibliography{references}

\newpage

\appendix

\section*{Appendix: The algorithm of Liu et al.}

The 2CCED algorithm of Liu et al. is claimed to have a worst-case running time in $\bigos(2.74^k)$ \cite{Liu12}. Unfortunately, there is a branching rule that is wrong due to an omitted case. 
The rule corresponds to the below figure (labeled Case 2.2.4 in the same paper). It is redrawn below for a clear illustration in a manner that matches our case analysis. 

\vspace{10pt}

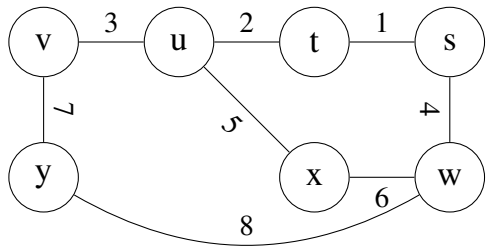
\begin{figure}[htb!]
    \centering
\begin{tikzpicture}[main/.style = {draw, circle, node distance=1.5cm, scale = 1.2, minimum size=2em},
every edge quotes/.style = {auto=left, sloped, font=\scriptsize, inner sep=1pt}
] 
\node[main] (1) {v};
\node[main] (2) [right of=1] {u};
\node[main] (3) [right of=2] {t};
\node[main] (4) [right of=3] {s};
\node[main] (5) [below of=3] {x};
\node[main] (6) [below of=4] {w};
\node[main] (7) [below of=1] {y};
\draw (1) -- node[above, sloped, pos=0.5] {3} (2);
\draw (1) -- node[above, sloped, pos=0.5] {7} (7);
\draw (2) -- node[above, sloped, pos=0.5] {2} (3);
\draw (3) -- node[above, sloped, pos=0.5] {1} (4);
\draw (2) -- node[below, sloped, pos=0.5] {5} (5);
\draw (4) -- node[below, sloped, pos=0.5] {4} (6);
\draw (6) -- node[below, sloped, pos=0.5] {6} (5);
\path (7) [bend right] edge node [above] {8} (6);

\end{tikzpicture}
    \caption{Case C.2.2.4 in \cite{Liu12}.}
    \label{4.2}
\end{figure}

\noindent
In \cite{Liu12}, the authors presented the following branching scenario (Page 245, Table 1, row 4).

\begin{itemize}
    \item[(1)] delete edges 1, 5 and 7;
    \item[(2)] delete edges 1, 5 and 8;
    \item[(3)] delete edges 1, 6 and 7;
    \item[(4)] delete edges 1, 6 and 8;
    \item[(5)] delete edges 2 and 4;
     \item[(6)] delete edges 2, 5 and 7;
     \item[(7)] delete edges 2, 5 and 8;
    \item[(8)] delete edges 2, 6 and 7;
     \item[(9)] delete edges 2, 6 and 8;
    \item[(10)] delete edges 3 and 7;
    \item[(11)] delete edges 3 and 8;
    \item[(12)] delete edges 3, 4 and 5;
    \item[(13)] delete edges 3, 4 and 6.
\end{itemize}

\noindent
The corresponding worst-case recurrence is $T(k) = 3T(k-2)+10T(k-3)$ with a running time in $\bigos(2.62^k)$. To understand the above branching, observe that it tries to resolve the conflict quadruple $(s,t,u,v)$ by first deleting edge 1 ($st$) and then simultaneously resolve the two conflict quadruples $(s,w,y,v)$ and $(s,w,x,u)$. The latter conflict results from the deletion of edge 1. 

The first four branches are not enough to cover the case of deleting edge 1 ($st$) since there is a case where both edges 1 and 4 are deleted. This becomes obvious from branches 5-9 where the authors do notice the need to delete edges 2 and 4 to cover the case where edge 2 is deleted. The branching rule can be fixed by adding a branch/case for the deletion of edges 1 and 4 at the beginning. The running time would go up to $\bigos(2.761^k)$ if this is fixed, provided there are no other errors or missed cases.  
Finally, had this branching rule been correct as described in \cite{Liu12}, we would have used it to cover Case 2.4 in our algorithm and we would have improved the running time to $\bigos(2.62^k)$.

\end{document}